\documentclass[10pt]{article}
\usepackage{palatino}
\usepackage{epsfig}
\usepackage{amsmath,amsfonts}
\usepackage{a4wide}

\newcommand{\half}{\mbox{\small{$\frac{1}{2}$}}}

\newcommand{\lms}{\Lambda_{\overline{\ensuremath{\mathrm{\tiny{MS}}}}}}
\newcommand{\Evac}{E_{\mbox{\tiny vac}}}
\newcommand{\MSbar}{\overline{\mbox{MS}}}

\newcommand{\occ}{\overline{c}}

\newcommand{\omu}{\overline{\mu}}
\newcommand{\p}{\partial}

\newcommand{\al}{\alpha}
\newcommand{\wsigma}{\widetilde{\sigma}}

\begin{document}
\date{}
\title{\textbf{Nonperturbative ghost dynamics in the maximal Abelian
gauge}}
\author{ \textbf{M.A.L. Capri$^a$\thanks{marcio@dft.if.uerj.br}} \ , \textbf{D. Dudal}$^{b}$\thanks{david.dudal@ugent.be}{\
} \ , \textbf{J.A.
Gracey$^{c}$\thanks{gracey@liv.ac.uk}} \ , \textbf{S.P. Sorella}$^{a}$\thanks{%
sorella@uerj.br}{\ }{\ }\ , \textbf{H. Verschelde}$^{b}$\thanks{henri.verschelde@ugent.be} \\\\
\textit{$^{a}$\small{Departamento de F\'{\i }sica Te\'{o}rica}}\\
\textit{\small{Instituto de F\'{\i }sica, UERJ, Universidade do Estado do Rio de Janeiro}} \\
\textit{\small{Rua S{\~a}o Francisco Xavier 524, 20550-013 Maracan{\~a}}} \\
\textit{\small{Rio de Janeiro, Brasil}} \\[3mm]
\textit{$^{b}$\small{Department of Mathematical Physics and Astronomy}} \\
\textit{\small{Ghent University}} \\
\textit{\small{Krijgslaan 281-S9, B-9000 Gent, Belgium}}\\
[3mm] \textit{$^c$\small{Theoretical Physics Division} }\\
\textit{\small{Department of Mathematical Sciences}}\\
\textit{\small{University of Liverpool}}\\
\textit{\small{P.O. Box 147, Liverpool, L69 3BX, United Kingdom}} }
\maketitle
\begin{abstract}
We construct the effective potential for the ghost condensate
$\langle f^{abi}\occ^a c^b\rangle$ in the maximal Abelian gauge.
This condensate is an order parameter for a global continuous
symmetry, which is spontaneously broken since a nonvanishing value
of $\langle f^{abi}\occ^a c^b\rangle$ lowers the vacuum energy. The
associated Goldstone mode turns out to be unphysical.
\end{abstract}
\vspace{-16cm} \hfill LTH--753 \vspace{15cm}

\section{Introduction}
Perturbatively, Faddeev-Popov ghosts are well understood. In
textbooks \cite{Peskin:1995ev}, these anticommuting scalar fields
are usually introduced as a tool to lift the Faddeev-Popov
determinant into the action. This determinant is the Jacobian
arising from the gauge fixing condition. After this, a consistent
perturbative expansion of the path integral can be carried out.

However, ghosts are much more than a ``mathematical trick''. In a
sense, they naturally arise when a gauge is fixed in a Lorentz
covariant manner. Once the gauge is fixed, the (local) gauge
freedom, generated by $\delta_\omega A_\mu^a =
D_{\mu}^{ab}\omega^b$, is of course lost. However, we recover a BRST
symmetry of the complete action $S_{\mbox{\tiny YM}}+S_{\mbox{\tiny
gf}}$, such that the gauge fixing part of the action is BRST exact,
{\it i.e.} $S_{\mbox{\tiny gf}}=sS_{\mbox{\tiny gf}}^\prime$, where
$s$ is the nilpotent BRST operator.

A more general class of gauge fixings than those which can be
obtained through the Faddeev-Popov method, can be introduced by
making direct use of the BRST symmetry: one adds a BRST-exact
expression $sS_{\mbox{\tiny gf}}^\prime$ to the classical Yang-Mills
action in order to break the gauge invariance.

Clearly, the ghosts play a crucial role in this construction. The
BRST symmetry can be used to prove e.g. the renormalizability of
gauge theories, the unitarity of the {\cal S}-matrix, the gauge
parameter independence of gauge invariant correlation functions,
etc.

Ghosts also have a clear physical meaning at the perturbative level,
although \emph{a clear unphysical meaning} would perhaps be a better
choice of words. Indeed, asymptotically, their degrees of freedom
cancel out with the scalar and longitudinal gauge boson
polarizations from a suitably defined physical subspace, leaving two
physical transverse polarizations, as desired
\cite{Henneaux:1992ig}.

Perhaps less commonly known is that ghosts can also be used to
discuss quantum properties of anomalies. For example, the ghost
polynomials $\mbox{Tr} c^{2k+1}$ can be used to discuss
nonrenormalization properties of e.g. the gauge anomaly
(Adler-Bardeen theorem). We refer to \cite{Piguet:1995er} for
relevant details and original literature.

This very short summary should have sufficiently outlined the
relevance of ghosts. The reader will have noticed that all the
previous results are strictly speaking at the perturbative level. A
natural question is what the role of the ghosts might become when
going beyond the perturbative level? Of course, since ghosts arise
only \emph{after} gauge fixing, one might question the issue of
gauge invariance. But as we have to add them to our action at the
quantum level, we cannot disregard the possibility that ghosts might
be important for the infrared dynamics of gauge theories, signalling
certain nonperturbative effects in at least a particular gauge. Let
us quote two cases in which ghosts are relevant at the
nonperturbative level. Firstly, when gauge copies in the Landau
gauge are taken into account by implementing the restriction of the
domain of integration in the functional integral to the so called
Gribov region $\Omega$ \cite{Gribov:1977wm}, the ghost propagator is
found to behave like $\frac{1}{q^4}$ for $q^2\sim 0$. At the same
time, an infrared suppressed gluon propagator is found. This leads
to an infrared fixed point of the nonperturbatively defined strong
coupling constant \cite{Gracey:2006dr}, and induces a violation of
positivity of the gluon propagator, indicating that the gluon cannot
be a stable asymptotic physical particle \cite{Dudal:2005na}.
Secondly, Kugo and Ojima constructed an algebraic criterion for
confinement (an inherent nonperturbative infrared phenomenon), which
is fulfilled in the Landau gauge when the ghost propagator is
sufficiently singular \cite{Kugo:1979gm,Kugo:1995km}. The infrared
enhancement of the ghost propagator and the infrared suppression of
the gluon propagator have received confirmation from lattice
simulations \cite{Maas:2006qw} as well as from the study of the
Schwinger-Dyson equations
\cite{Watson:2001yv,Lerche:2002ep,Zwanziger:2001kw}.

In this paper, we will elaborate on another intriguing possibility:
the formation of a nonperturbative ghost condensate. This problem
was tackled first in \cite{Schaden:1999ew,Kondo:2000ey} in case of
the maximal Abelian gauge (MAG). The MAG Yang-Mills action is given
by\footnote{Indices like $a,b,\ldots$ refer to the off-diagonal
sector, while $i,j,\ldots$ to the diagonal one.}
\begin{eqnarray}\label{totaal}
S=S_{\mbox{\tiny YM}}+S_{\mbox{\tiny MAG}}+S_{\mbox{\tiny diag}}\;,
\end{eqnarray}
whereby
\begin{eqnarray}
S_{\mbox{\tiny YM}}\!\!\!\! &=&\!\!\!\!-\frac{1}{4}\int {d^{4}\!x}\,\Bigl%
(F_{\mu \nu }^{a}F^{a\mu \nu }+F_{\mu \nu }^{i}F^{i\mu \nu
}\Bigr)\;,
\label{YM} \\
&&  \nonumber \\
S_{\mbox{\tiny MAG}}\!\!\!\! &=&\!\!\!\!s\overline{s}\int {d^{4}\!x}\,\biggl(%
\frac{1}{2}A_{\mu }^{a}A^{a\mu }-\frac{\alpha
}{2}c^{a}\overline{c}^{a}\biggl)
\nonumber \\
&=&\!\!\!\!\int {d^{4}\!x}\,\biggl[b^{a}\biggl(D_{\mu }^{ab}A^{b\mu }+\frac{%
\alpha }{2}b^{a}\biggl)+\overline{c}^{a}D_{\mu }^{ab}D^{bc\mu }c^{c}+gf^{abi}\overline{%
c}^{a}(D_{\mu }^{bc}A^{c\mu })c^{i}+gf^{bcd}\overline{c}^{a}D_{\mu
}^{ab}(A^{c\mu
}c^{d})  \nonumber \\
&&\!\!\!\!-\alpha gf^{abi}b^{a}\overline{c}^{b}c^{i}-g^{2}f^{abi}f^{cdi}\overline{c}%
^{a}c^{d}A_{\mu }^{b}A^{c\mu }-\frac{\alpha }{2}gf^{abc}b^{a}\overline{c}%
^{b}c^{c}-\frac{\alpha }{4}g^{2}f^{abi}f^{cdi}\overline{c}^{a}\overline{c}%
^{b}c^{c}c^{d}  \nonumber \\
&&\!\!\!\!-\frac{\alpha }{4}g^{2}f^{abc}f^{adi}\overline{c}^{b}\overline{c}%
^{c}c^{d}c^{i}-\frac{\alpha }{8}g^{2}f^{abc}f^{ade}\overline{c}^{b}\overline{c}%
^{c}c^{d}c^{e}\biggr]\;,  \label{MAG} \\
&&  \nonumber \\
S_{\mbox{\tiny diag}}\!\!\!\! &=&\!\!\!\!s\int {d^{4}\!x}\,\overline{c}%
^{i}\partial ^{\mu }\!A_{\mu }^{i}=\int
{d^{4}\!x}\,\Bigl[b^{i}\partial ^{\mu }\!A_{\mu
}^{i}+\overline{c}^{i}\partial ^{\mu }(\partial _{\mu
}c^{i}+gf^{abi}A_{\mu }^{a}c^{b})\Bigr]\;,  \label{diag}
\end{eqnarray}
with
\begin{equation}
D_{\mu }^{ab}=\delta ^{ab}\partial _{\mu }-gf^{abi}A_{\mu }^{i}
\label{cov_derivative}
\end{equation}
the $U(1)^{N-1}$ covariant derivative and
\begin{eqnarray}
F_{\mu \nu }^{a}\!\!\!\! &=&\!\!\!\!D_{\mu }^{ab}A_{\nu }^{b}-D_{\nu
}^{ab}A_{\mu }^{b}+gf^{abc}A_{\mu }^{b}A_{\nu }^{c}\;, F_{\mu \nu
}^{i}=\partial _{\mu }A_{\nu }^{i}-\partial _{\nu }A_{\mu
}^{i}+gf^{abi}A_{\mu }^{a}A_{\nu }^{b}  \label{F}
\end{eqnarray}
the field strength. The nilpotent (anti)-BRST transformations of
the fields read as follows
\begin{eqnarray}
sA_{\mu }^{a}\!\!\!\! &=&\!\!\!\!-(D_{\mu }^{ab}c^{b}+gf^{abc}A_{\mu
}^{b}c^{c}+gf^{abi}A_{\mu }^{b}c^{i})\;,  sA_{\mu }^{i}=-(\partial
_{\mu }c^{i}+gf^{abi}A_{\mu
}^{a}c^{b})\;,  \nonumber \\
sc^{a}\!\!\!\!
&=&\!\!\!\!gf^{abi}c^{b}c^{i}+\frac{g}{2}f^{abc}c^{b}c^{c}\;,sc^{i}=\frac{g}{2}f^{abi}c^{a}c^{b}\;,  \nonumber \\
s\overline{c}^{a}\!\!\!\!
&=&\!\!\!\!b^{a}\;,sb^{a}=0\;,s\overline{c}^{i}=b^{i}\;,sb^{i}=0\;,
\label{BRST}
\end{eqnarray}
and
\begin{eqnarray}
\overline{s}A_{\mu }^{a}\!\!\!\! &=&\!\!\!\!-(D_{\mu }^{ab}\overline{c}%
^{b}+gf^{abc}A_{\mu }^{b}\overline{c}^{c}+gf^{abi}A_{\mu
}^{b}\overline{c}^{i})\;,\overline{s}A_{\mu }^{i}=-(\partial _{\mu }\overline{c}%
^{i}+gf^{abi}A_{\mu }^{a}\overline{c}^{b})\;,  \nonumber \\
\overline{s}\overline{c}^{a}\!\!\!\! &=&\!\!\!\!gf^{abi}\overline{c}^{b}\overline{c}^{i}+\frac{g%
}{2}f^{abc}\overline{c}^{b}\overline{c}^{c}\;,  \overline{s}\overline{c}^{i}=\frac{g}{2}f^{abi}\overline{c}^{a}\overline{c}%
^{b}\;,  \nonumber \\
\overline{s}c^{a}\!\!\!\! &=&\!\!\!\!-b^{a}+gf^{abc}c^{b}\overline{c}^{c}+gf^{abi}c^{b}%
\overline{c}^{i}+gf^{abi}\overline{c}^{b}c^{i}\;,  \nonumber \\
\overline{s}c^{i}\!\!\!\!
&=&\!\!\!\!-b^{i}+gf^{abi}c^{a}\overline{c}^{b}\;, \overline{s}b^{a}=-gf^{abc}b^{b}\overline{c}^{c}-gf^{abi}b^{b}\overline{c%
}^{i}+gf^{abi}\overline{c}^{b}b^{i}\;\,
\overline{s}b^{i}=-gf^{abi}b^{a}\overline{c}^{b}\;.
\label{anti-BRST}
\end{eqnarray}
We also recall the Jacobi identity, in decomposed form
\cite{Dudal:2004rx}
\begin{equation}\label{jac}
    f^{abi}f^{bjc}+f^{abj}f^{bci}=0\;,\qquad
    f^{abc}f^{bdi}+f^{abd}f^{bic}+f^{abi}f^{bcd}=0\;.
\end{equation}
Strictly speaking, the MAG is defined by choosing that gauge
configuration that corresponds to the (absolute) minimum of the
functional
\begin{equation}\label{MAG1}
    \mathcal{R}_{\mbox{\tiny MAG}}[A]=\int d^4x (A_\mu^a)^2
\end{equation}
under gauge variations. Restricting to infinitesimal gauge
variations, it reduces to the $U(1)^{N-1}$ covariant constraint
\begin{equation}\label{MAG2}
    D_\mu^{ab}A^{b\mu}=0\;.
\end{equation}
The residual Abelian gauge freedom is fixed by a Landau like
condition, \eqref{diag}. We notice that this gauge fixing for the
diagonal part does not exhibit the anti-BRST symmetry. We have only
introduced the anti-BRST transformation $\overline{s}$ as a tool to
write down a condensed form of the off-diagonal gauge fixing, i.e.
the MAG \eqref{MAG}. As a consequence, only the BRST symmetry and
its associated Slavnov-Taylor identity will be used for the
renormalization analysis.

The MAG has received much interest, as it might be relevant for the
dual superconductivity picture of confinement \cite{'tHooft:1981ht}.
To make the MAG well-defined at the perturbative level, one must
introduce a regulating gauge parameter $\al$ and add a 4-point ghost
interaction proportional to $\al$ to the action
\cite{Min:1985bx,Fazio:2001rm}. The condition \eqref{MAG2} is
retrieved in the formal limit $\alpha\to 0$. As is well known from
other models, in the presence of an attractive four fermion
interaction, the formation of a fermion condensate can become
energetically favoured. In our case, the analog phenomenon would be
the formation of a ghost condensate. This was originally discussed
in \cite{Schaden:1999ew,Kondo:2000ey} by a decomposition of the
4-point interaction by means of an auxiliary field $\sigma$. A one
loop effective potential was constructed, and a nonvanishing
condensate $\langle\sigma\rangle$, proportional to the ghost
condensate, was found. However, as explained in \cite{Dudal:2002xe},
this gives rise to problems with the renormalization group (RG)
beyond the one loop level. Next to this, it is also of no use in the
Landau gauge as there is no 4-point interaction present in that
case. In this article, we shall invoke the LCO formalism, originated
in \cite{Verschelde:1995jj} by one of us, which allows a RG
consistent discussion of Local Composite Operators. We shall also
make use of results obtained in a series of papers about the usage
of the LCO formalism in gauge theories
\cite{Dudal:2003by,Dudal:2004rx}. The ghost condensate was used in
\cite{Schaden:1999ew,Kondo:2000ey} to generate a dynamical
off-diagonal gluon mass. As a consequence, the off-diagonal gluons
should decouple from the infrared dynamics, hinting that an Abelian
theory could be used to eventually obtain confinement. It was
however realized in \cite{Dudal:2002xe} that the effective
off-diagonal gluon mass was tachyonic, and therefore certainly not
suitable to explain the so-called Abelian dominance
\cite{Ezawa:1982bf}.

Nevertheless, this does not mean that the ghost condensate is
meaningless. The fact that it gives rise to a tachyonic effective
gluon mass points out that other condensates might emerge. Indeed,
it was discussed in \cite{Dudal:2004rx,Kondo:2001nq} that a mixed
gluon-ghost operator condenses and gives rise to a real valued
effective off-diagonal mass, a result in qualitative accordance with
available lattice simulations in the MAG
\cite{Amemiya:1998jz,Bornyakov:2003ee}.

In the Landau gauge, we already presented a combined study of the
gauge condensate $\langle A^2\rangle$ together with the ghost
condensate \cite{Capri:2005vw}. The one loop net result is that the
tachyonic contribution of the ghost condensate induces a splitting
between the diagonal and off-diagonal mass, leaving a larger value
for the off-diagonal one. This can be seen as evidence for some kind
of Abelian dominance in the Landau gauge \cite{Suzuki:2004dw}.

A nontrivial condensation is frequently intimately entangled with
the spontaneous breaking of some global symmetry. In fact, as
discussed in \cite{Schaden:1999ew,Kondo:2000ey,Dudal:2003by}, the
ghost condensation breaks a global invariance present in the maximal
Abelian gauge, generated by
\begin{eqnarray}\label{symgen}
  \delta \occ^a &=& c^a\;,\;\;\;\;\delta b^a=\frac{g}{2}f^{abc}c^bc^c+gf^{abi}c^bc^i\;,
  \;\;\;\;\delta(\mbox{rest})=0\;.
\end{eqnarray}
As a version of this symmetry is also present in the Curci-Ferrari ,
see \cite{Curci:1976bt}, and Landau gauges \cite{Dudal:2002ye}, it
might be expected that a ghost condensation could occur in these
gauges too. This point was discussed in
\cite{Lemes:2002jv,Lemes:2002rc,Dudal:2003dp}. By choosing another
diagonal gauge fixing in the case of the MAG \cite{Dudal:2002ye}, it
is possible to find an even larger symmetry content, next to
$\delta$. More precisely, it is possible to find a complete
$SL(2,\mathbb{R})$ invariance, generated by the ghost number
symmetry generator $\delta_c$, $\delta$ and an analogue of $\delta$
with an exchange of ghost-antighost fields. In the Landau gauge, the
$SL(2,\mathbb{R})$ rotations connect different channels of ghost
condensation. Next to the operator $f^{abc}\occ^b c^c$ with
vanishing ghost number, also the ghost charged operators
$f^{abc}\occ^b \occ^c$ and $f^{abc}c^b c^c$ can condense in
principle. However, the corresponding vacua are equivalent
\cite{Dudal:2003dp}, and for simplicity we will restrict ourselves
in this paper to the uncharged channel $f^{abi}\occ^a c^b$ instead
of the charged ones ($f^{abi}c^a c^b/f^{abi}\occ^a \occ^b$). In
\cite{Dudal:2003by}, both channels were discussed simultaneously in
the Landau gauge.

Let us finally mention that, precursored by the theoretical results,
also lattice studies have been made in the Landau as well as in the
maximal Abelian gauge, giving support to the existence of a
nonvanishing ghost condensate and dynamically broken symmetry
\cite{Cucchieri:2005yr,Mendes:2006kc}.

This article is organized as follows: in the second section, we set
up the action and prove the renormalizability in the presence of the
composite ghost operator $f^{abi}\occ^a c^b$ and of the dimension
two mass operator $(\frac{1}{2}A^a_\mu A^a_\mu +\alpha \occ^a c^a)$
using the Ward identities of the MAG. In the third section, we
derive the necessary RG functions and discuss the one loop ghost
condensation for the gauge group $SU(2)$. Section 4 summarizes some
consequences of a nontrivial ghost condensate. We also prove that
the Goldstone boson corresponding to the spontaneously broken
$\delta$-symmetry decouples from the physical spectrum. The last
section contains our conclusions.

\section{The action: algebraic analysis}
The complete action we start with, reads
\begin{equation}
\Sigma =S_{\mbox{\tiny YM}}+S_{\mbox{\tiny MAG}}+S_{\mbox{\tiny diag}}+S_{%
\mbox{\tiny LCO}}+S_{\mbox{\tiny ext}}\;,  \label{action}
\end{equation}
\noindent where,
\begin{eqnarray}
S_{\mbox{\tiny LCO}}\!\!\!\! &=&\!\!\!\!s\int {d^{4}\!x}\,\biggl[\lambda %
\biggl(\frac{1}{2}A_{\mu }^{a}A^{a\mu }+\alpha \overline{c}^{a}c^{a}\biggr)+\frac{%
\zeta }{2}\lambda J+gf^{abi}\omega ^{i}\overline{c}^{a}c^{b}+\frac{\chi }{2}%
\omega ^{i}\vartheta ^{i}\biggr]  \nonumber \\
&=&\!\!\!\!\int {d^{4}\!x}\,\biggl[J\biggl(\frac{1}{2}A_{\mu
}^{a}A^{a\mu }+\alpha \overline{c}^{a}c^{a}\biggr)+\frac{\zeta
}{2}J^{2}-\alpha \lambda
b^{a}c^{a}+\lambda A^{a\mu }D_{\mu }^{ab}c^{b}+\alpha \lambda gf^{abi}\overline{c}%
^{a}c^{b}c^{i}  \nonumber \\
&&\!\!\!\!+\alpha \lambda \frac{g}{2}f^{abc}\overline{c}^{a}c^{b}c^{c}+gf^{abi}%
\vartheta ^{i}\overline{c}^{a}c^{b}-gf^{abi}\omega
^{i}b^{a}c^{b}+g^{2}f^{abi}f^{bcj}\omega
^{i}\overline{c}^{a}c^{c}c^{j} \nonumber
\\
&&\!\!\!+\frac{g^{2}}{2}f^{abi}f^{bcd}\omega ^{i}\overline{c}^{a}c^{c}c^{d}+\frac{%
\chi }{2}\vartheta ^{i}\vartheta ^{i}\biggr]\;,  \label{LCO} \\
&&  \nonumber \\
S_{\mbox{\tiny ext}}\!\!\!\! &=&\!\!\!\!\int {d^{4}\!x}\,\biggl[-\Omega
^{a\mu }\Bigl(D_{\mu }^{ab}c^{b}+gf^{abc}A_{\mu }^{b}c^{c}+gf^{abi}A_{\mu
}^{b}c^{i}\Bigr)-\Omega ^{i\mu }\Bigl(\partial _{\mu }c^{i}+gf^{abi}A_{\mu
}^{a}c^{b}\Bigr)  \nonumber \\
&&\!\!\!\!+L^{a}\biggl(gf^{abi}c^{b}c^{i}+\frac{g}{2}f^{abc}c^{b}c^{c}\biggr)%
+\frac{g}{2}f^{abi}L^{i}c^{a}c^{b}\biggr]\;.  \label{ext}
\end{eqnarray}
The external sources $\Omega _{\mu }^{a,i}$ and $L^{a,i}$ are needed
to define the composite operators entering the nonlinear BRST
transformations of the field $A_{\mu }^{a,i}$ and $c^{a,i}$,
respectively. These sources are invariant under the action of the
BRST operator, i.e.,
\begin{eqnarray}
&&s\Omega _{\mu }^{a}=s\Omega _{\mu }^{i}=0\;,\qquad
sL^{a}=sL^{i}=0.
\end{eqnarray}
\noindent The two pairs of sources $(J,\lambda )$ and $(\vartheta
^{i},\omega ^{i})$ are needed in order to define the composite operators $%
\left( A_{\mu }^{a}A^{a\mu }+\alpha \overline{c}^{a}c^{c}\right) $ and  $gf^{abi}%
\overline{c}^{a}c^{b}$ and their BRST variations. These sources form
BRST\ doublets, according to
\begin{eqnarray}
&&s\lambda =J,\qquad sJ=0\;, \qquad s\omega ^{i}=\vartheta
^{i},\qquad s\vartheta ^{i}=0.
\end{eqnarray}
The purpose of the pure source terms $\frac{\chi }{2}%
\vartheta ^{i}\vartheta ^{i}$ and $\frac{\zeta }{2}J^2$ shall be
made clear in the next section. The mass dimension and the ghost
number of the fields and sources have been listed in Table
\ref{table1}.
\begin{table}[t]
\centering
\begin{tabular}{|c|c|c|c|c|c|c|c|c|c|c|}
\hline & $A^{a,i}_{\mu}$ & $c^{a,i}$ & $\overline{c}^{a,i}$ &
$b^{a,i}$ & $\lambda$ & $J$ & $\Omega^{a,i}_{\mu}$ & $L^{a,i}$ &
$\omega^{i}$ & $\vartheta^{i}$ \\ \hline
dimension & 1 & 0 & 2 & 2 & 2 & 2 & 3 & 4 & 2 & 2 \\
ghost number & 0 & 1 & $-1$ & 0 & $-1$ & 0 & $-1$ & $-2$ & $-1$ & 0 \\ \hline
\end{tabular}
\caption{Quantum numbers of the field and sources}
\label{table1}
\end{table}
The complete action (\ref{action}) obeys  the following set of Ward
identities:
\begin{itemize}
\item  The Slavnov-Taylor identity
\begin{eqnarray}
\mathcal{S}(\Sigma )\!\!\!\! &=&\!\!\!\!\int {d^{4}\!x}\,\biggl(\frac{\delta
\Sigma }{\delta \Omega ^{a\mu }}\frac{\delta \Sigma }{\delta A_{\mu }^{a}}+%
\frac{\delta \Sigma }{\delta \Omega ^{i\mu }}\frac{\delta \Sigma }{\delta
A_{\mu }^{i}}+\frac{\delta \Sigma }{\delta L^{a}}\frac{\delta \Sigma }{%
\delta c^{a}}+\frac{\delta \Sigma }{\delta L^{i}}\frac{\delta \Sigma }{%
\delta c^{i}}+b^{a}\frac{\delta \Sigma }{\delta \overline{c}^{a}}  \nonumber \\
&&\!\!\!\!+b^{i}\frac{\delta \Sigma }{\delta \overline{c}^{i}}+\vartheta ^{i}%
\frac{\delta \Sigma }{\delta \omega ^{i}}+J\frac{\delta \Sigma }{\delta
\lambda }\biggr)=0\;,  \label{ST}
\end{eqnarray}
\item  The diagonal ghost equation
\begin{eqnarray}
&&\!\!\!\!\mathcal{G}^{i}(\Sigma )=\Delta _{\mbox{\tiny class}}^{i}\;,
\nonumber \\
&&\!\!\!\!\mathcal{G}^{i}=\frac{\delta }{\delta c^{i}}+gf^{abi}\overline{c}^{a}%
\frac{\delta }{\delta b^{b}}\;,  \nonumber \\
&&\!\!\!\!\Delta _{\mbox{\tiny class}}^{i}=-\partial ^{2}\overline{c}%
^{i}+gf^{abi}\Omega ^{a\mu }A_{\mu }^{b}-\partial _{\mu }\Omega ^{i\mu
}-gf^{abi}L^{a}c^{b}\;.  \label{ghost-eq}
\end{eqnarray}
\item  The diagonal gauge-fixing condition
\begin{equation}
\frac{\delta \Sigma }{\delta b^{i}}=\partial _{\mu }A^{i\mu }\;.  \label{gf}
\end{equation}
\item  The anti-ghost equation
\begin{equation}
\overline{\mathcal{G}}^{i}(\Sigma )=\frac{\delta \Sigma }{\delta \overline{c}^{i}}%
+\partial ^{\mu }\frac{\delta \Sigma }{\delta \Omega ^{i\mu }}=0\;.
\label{anti-gh-eq}
\end{equation}
\item  The diagonal $U(1)^{N-1}$ Ward identity
\begin{eqnarray}
&&\hspace{-30pt}\mathcal{W}^{i}(\Sigma )=-\partial ^{2}b^{i}\;,  \nonumber \\
&&\hspace{-30pt}\mathcal{W}^{i}=\partial _{\mu }\frac{\delta }{\delta A_{\mu
}^{i}}+gf^{abi}\biggl(A_{\mu }^{a}\frac{\delta }{\delta A_{\mu }^{b}}+c^{a}%
\frac{\delta }{\delta c^{b}}+b^{a}\frac{\delta }{\delta b^{b}}+\overline{c}^{a}%
\frac{\delta }{\delta \overline{c}^{b}}+\Omega ^{a\mu }\frac{\delta
}{\delta \Omega ^{b\mu }}+L^{a}\frac{\delta }{\delta
L^{b}}\biggr)\;.  \label{U(1)-eq}
\end{eqnarray}
\item  The integrated $\lambda $-equation
\begin{equation}
\mathcal{U}(\Sigma )=\int {d^{4}\!x}\,\biggl(\frac{\delta \Sigma }{\delta
\lambda }+c^{a}\frac{\delta \Sigma }{\delta b^{a}}-2\omega ^{i}\frac{\delta
\Sigma }{\delta L^{i}}\biggr)=0\;.  \label{lambda-eq}
\end{equation}
\item  The $SL(2,\mathbb{R})$ Ward identity
\begin{equation}
D(\Sigma )=\int {d^{4}\!x}\,\biggl(c^{a}\frac{\delta \Sigma }{\delta \overline{c}%
^{a}}+\frac{\delta \Sigma }{\delta L^{a}}\frac{\delta \Sigma }{\delta b^{a}}%
-2\vartheta ^{i}\frac{\delta \Sigma }{\delta L^{i}}\biggr)=0\;.  \label{sl2r}
\end{equation}
\end{itemize}
We notice that the terms $\Delta _{\mbox{\tiny class}}^{i}$, in
(\ref{ghost-eq}), and $-\partial ^{2}b^{i}$, in (\ref{U(1)-eq}), are
linear in the quantum fields, thus defining classical breakings.

We are now ready to write down the most general counterterm, $\Sigma
_{\mbox{\tiny CT}}$, which is compatible with the previous Ward
identities and which can be freely added to the original action.
Requiring
that the perturbed action, $\Sigma +\eta \Sigma _{\mbox{\tiny CT}}$%
, obeys the same Ward identities as $\Sigma $ to the first order in
the expansion parameter $\eta $, one gets the following conditions
\begin{eqnarray}
\mathcal{B}_{\Sigma }\Sigma _{\mbox{\tiny CT}}\!\!\!\!
&=&\!\!\!\!0\;, \frac{\delta \Sigma _{\mbox{\tiny CT}}}{\delta
b^{i}}=0\;, \mathcal{G}^{i}\Sigma _{\mbox{\tiny CT}}=0\;,
\mathcal{W}^{i}\Sigma _{\mbox{\tiny CT}}=0\;,
\overline{\mathcal{G}}^{i}\Sigma _{\mbox{\tiny CT}}=0,
\mathcal{U}\Sigma _{\mbox{\tiny CT}}=0\;, D_{\Sigma }\Sigma
_{\mbox{\tiny CT}}=0\;,  \label{D-eq}
\end{eqnarray}
where $\mathcal{B}_{\Sigma }$ is the nilpotent, $\mathcal{B}_{\Sigma
}^{2}=0$, linearized Slavnov-Taylor operator, given by
\begin{eqnarray}
\mathcal{B}_{\Sigma }\!\!\!\! &=&\!\!\!\!\int {d^{4}\!x}\,\biggl(\frac{%
\delta \Sigma }{\delta \Omega ^{a\mu }}\frac{\delta }{\delta A_{\mu }^{a}}+%
\frac{\delta \Sigma }{\delta A_{\mu }^{a}}\frac{\delta }{\delta \Omega
^{a\mu }}+\frac{\delta \Sigma }{\delta \Omega ^{i\mu }}\frac{\delta }{\delta
A_{\mu }^{i}}+\frac{\delta \Sigma }{\delta A_{\mu }^{i}}\frac{\delta }{%
\delta \Omega ^{i\mu }}+\frac{\delta \Sigma }{\delta L^{a}}\frac{\delta }{%
\delta c^{a}}+  \nonumber \\
&&\!\!\!\!+\frac{\delta \Sigma }{\delta c^{a}}\frac{\delta }{\delta L^{a}}+%
\frac{\delta \Sigma }{\delta L^{i}}\frac{\delta }{\delta c^{i}}+\frac{\delta
\Sigma }{\delta c^{i}}\frac{\delta }{\delta L^{i}}+b^{a}\frac{\delta }{%
\delta \overline{c}^{a}}+b^{i}\frac{\delta }{\delta \overline{c}^{i}}+\vartheta ^{i}%
\frac{\delta }{\delta \omega ^{i}}+J\frac{\delta }{\delta \lambda
}\biggr)\;,   \label{lp}
\end{eqnarray}
and the operator $D_{\Sigma }$, in (\ref{D-eq}), is given by:
\begin{eqnarray}
D_{\Sigma }&=&\int {d^{4}\!x}\,\biggl(c^{a}\frac{\delta }{%
\delta \overline{c}^{a}}+\frac{\delta \Sigma }{\delta
L^{a}}\frac{\delta }{\delta
b^{a}}+\frac{\delta \Sigma }{\delta b^{a}}\frac{\delta }{\delta L^{a}}%
-2\vartheta ^{i}\frac{\delta }{\delta L^{i}}\biggr)\;.
\end{eqnarray}
The most general local counterterm can be
written as
\begin{equation}
\Sigma _{\mbox{\tiny CT}}=a_{0}\,S_{\mbox{\tiny YM}}+\mathcal{B}_{\Sigma
}\Delta ^{(-1)}\;,  \label{general-CT}
\end{equation}
where $\Delta ^{(-1)}$ is an integrated local polynomial of ghost
number $-1$ and dimension 4, given by:
\begin{eqnarray}
\Delta ^{(-1)}\!\!\!\! &=&\!\!\!\!\int
{d^{4}\!x}\,\biggl[a_{1}\,\Omega ^{i\mu }A_{\mu
}^{i}+a_{2}\,(\partial ^{\mu }\overline{c}^{i})A_{\mu
}^{i}+a_{3}\,\Omega ^{a\mu }A_{\mu }^{a}+a_{4}\,(\partial ^{\mu }\overline{c}%
^{a})A_{\mu }^{a}+a_{5}\,L^{a}c^{a}  \nonumber \\
&&\!\!\!\!+a_{6}\,L^{i}c^{i}+a_{7}(\partial ^{\mu }\!A_{\mu
}^{i})\omega ^{i}+a_{8}\,\lambda c^{i}\omega ^{i}+a_{9}\,\omega
^{i}\vartheta ^{i}+a_{10}\,gf^{abi}\omega
^{i}\overline{c}^{a}c^{b}+a_{11}\,\lambda J  \nonumber
\\
&&\!\!\!\!+a_{12}\,\lambda A_{\mu }^{a}A^{a\mu }+a_{13}\,\lambda A_{\mu
}^{i}A^{i\mu }+a_{14}\,\lambda \overline{c}^{a}c^{a}+a_{15}\,gf^{abc}\overline{c}^{a}%
\overline{c}^{b}c^{c}+a_{16}\,gf^{abi}\overline{c}^{c}\overline{c}^{b}c^{i}  \nonumber \\
&&\!\!\!\!+a_{17}\,\overline{c}^{a}b^{a}+a_{18}\,gf^{abi}\overline{c}^{a}A_{\mu
}^{b}A^{i\mu }+a_{19}\,gf^{abi}\overline{c}^{i}\overline{c}^{a}c^{b}+a_{20}\,b^{i}\overline{%
c}^{i}+a_{21}\,\lambda \overline{c}^{i}c^{i}  \nonumber \\
&&\!\!\!\!+a_{22}\,\overline{c}^{i}\vartheta
^{i}+a_{23}\,b^{i}\omega ^{i}\biggr] \label{Delta-1}
\end{eqnarray}
The identities \eqref{D-eq} imply that
\begin{eqnarray}
&&a_{1}=a_{2}=a_{6}=a_{7}=a_{8}=a_{19}=a_{20}=a_{21}=a_{22}=a_{23}=0\;, \nonumber\\
&&a_{10}=-a_{5}\;, a_{12}=-\frac{a_{4}}{2}+\frac{a_{5}}{2}\;,
a_{14}=-2a_{16}+\alpha a_{5}\;, a_{15}=\frac{a_{16}}{2}\;,
a_{17}=-a_{16}\;, a_{18}=a_{4}\;.
\end{eqnarray}
If we rename the six independent coefficients $a_{3}$, $a_{4}$, $%
a_{5}$, $a_{9}$, $a_{11}$, $a_{16}$, according to
\begin{eqnarray}
a_{3} &\to &{a_{1}},\quad {a_{4}}\to -{a_{3}},\quad {a_{5}}\to {a_{2}},\quad
{a_{9}}\to \frac{{a_{5}}\chi }{2},\quad {a_{11}}\to \frac{{a_{6}}\zeta }{2}%
,\quad {a_{16}}\to -\alpha {a_{4}}\;,  \label{rename}
\end{eqnarray}
the final expression for $\Delta ^{(-1)}$ is found to be
\begin{eqnarray}
\Delta ^{(-1)}\!\!\!\! &=&\!\!\!\!\int {d^{4}\!x}\,\biggl[a_{1}\,\Omega
^{a\mu }A_{\mu }^{a}+a_{2}\,\biggl(L^{a}c^{a}-gf^{abi}\omega ^{i}\overline{c}%
^{a}c^{b}+\frac{1}{2}\lambda A_{\mu }^{a}A^{a\mu }+\alpha \lambda \overline{c}%
^{a}c^{a}\biggr)  \nonumber \\
&&\!\!\!\!+a_{3}\,\biggl(\overline{c}^{a}D_{\mu }^{ab}A^{b\mu }+\frac{1}{2}%
\lambda A_{\mu }^{a}A^{a\mu }\biggr)+\alpha a_{4}\,\biggl(\overline{c}%
^{a}b^{a}-gf^{abi}\overline{c}^{a}\overline{c}^{b}c^{i}-\frac{g}{2}f^{abc}\overline{c}^{a}%
\overline{c}^{b}c^{c}  \nonumber \\
&&\!\!\!\!+2\lambda \overline{c}^{a}c^{a}\biggr)+\frac{a_{5}\chi
}{2}\,\omega ^{i}\vartheta ^{i}+\frac{a_{6}\zeta }{2}\lambda
J\biggr]\;. \label{final-Delta}
\end{eqnarray}
At the end, $\Sigma _{\mbox{\tiny CT}}$, in (\ref{general-CT}%
), contains seven free independent parameters $a_{k}$ $(k=0,1,\dots ,6)$.
These parameters can be reabsorbed by means of a multiplicative
renormalization of the parameters $\xi =(g,\alpha ,\zeta ,\chi )$, of the
fields $\Phi =(A_{\mu }^{a,i},c^{a,i},\overline{c}^{a,i},b^{a,i})$ and sources $%
\phi =(\Omega _{\mu }^{a,i},L^{a,i},\lambda ,J,\omega ^{i},\vartheta ^{i})$,
according to
\begin{equation}
\Sigma (\Phi _{0},\phi _{0},\xi _{0})=\Sigma (\Phi ,\phi ,\xi )+\eta
\Sigma _{\mbox{\tiny CT}}(\Phi ,\phi ,\xi )\;,
\end{equation}
where,
\begin{eqnarray}
\Phi _{0}^{\mbox{\tiny diag}}\!\!\!\! &=&\!\!\!\!Z_{\Phi }^{1/2}\Phi ^{%
\mbox{\tiny diag}}\;, \Phi _{0}^{\mbox{\tiny
off-diag}}={\widetilde{Z}}_{\Phi
}^{1/2}\Phi ^{\mbox{\tiny off-diag}}\;, \\
\phi _{0}^{\mbox{\tiny diag}}\!\!\!\! &=&\!\!\!\!Z_{\phi }\phi
^{\mbox{\tiny diag}}\;,  \phi _{0}^{\mbox{\tiny
off-diag}}={\widetilde{Z}}_{\phi
}\phi ^{\mbox{\tiny off-diag}}\;, \\
\xi _{0}\!\!\!\! &=&\!\!\!\!Z_{\xi }\xi .  \label{renormalizations}
\end{eqnarray}
More precisely, a little algebra results in
\begin{eqnarray}
Z_{g}\!\!\!\! &=&\!\!\!\!1-\eta \,\frac{a_{0}}{2}\;,
\widetilde{Z}_{A}=1+\eta \,(a_{0}+2a_{1})\;, Z_{c}=1+\eta
\,(a_{2}+a_{3})\;,
\widetilde{Z}_{c}=1-\eta \,(a_{2}-a_{3})\;, \nonumber\\
Z_{\alpha }\!\!\!\! &=&\!\!\!\!1+\eta \,(a_{0}-2a_{3}+2a_{4})\;,
Z_{\chi }=1-\eta(a_0-2a_2-2a_3-a_5) \;, Z_{\zeta }=1+\eta
\,(2a_{0}-2a_{2}-2a_{3}+a_{6})\;,\nonumber\\
\end{eqnarray}
and
\begin{eqnarray}
Z_{A}\!\!\!\! &=&\!\!\!\!Z_{g}^{-2}\;,
Z_{b}=Z_{g}^{2}\;,\widetilde{Z}_{b}=Z_{g}^{2}Z_{c}\widetilde{Z}_{c}\;,
Z_{\overline{c}}=Z_{c}^{-1}\;,
\widetilde{Z}_{\overline{c}}=\widetilde{Z}_{c}\;,Z_{\Omega
}=Z_{c}^{-1/2}\;, \widetilde{Z}_{\Omega }=Z_{g}^{-1}\widetilde{Z}%
_{A}^{-1/2}Z_{c}^{-1/2}\;, \nonumber\\
Z_{L}\!\!\!\! &=&\!\!\!\!Z_{g}^{-1}Z_{c}^{-1}\;, \widetilde{Z}_{L}=Z_{g}^{-1}Z_{c}^{-1/2}\widetilde{Z}%
_{c}^{-1/2}\;, Z_{\lambda }=Z_{g}Z_{c}^{1/2}\;,
Z_{J}=Z_{g}^{2}Z_{c}\;Z_{\omega }=Z_{g}^{-2}Z_{c}^{-3/2}\;,
Z_{\vartheta }= Z_g^{-1}Z_c^{-1} \nonumber\\\label{Ztheta}
\end{eqnarray}
Before closing this section, we notice that there is no mixing at
all between the mass operator coupled to $J$ and the ghost operator
coupled to $\vartheta^i$.

\section{Construction of the effective potential}
We shall employ dimensional regularization in $d=4-\varepsilon$
dimensions. The part of the action \eqref{action} that we need is
obtained by setting all external sources equal to zero, except
$\vartheta^i$ which is coupled to the operator $gf^{abi}\occ^ac^b$.
For the moment, we also discard the mass operator, and concentrate
purely on the dimension two ghost operator. For further analysis, we
prefer to use the operator $f^{abi}\occ^ac^b$, obtained by a
suitable rescaling of the original operator coupled to the source
$\vartheta^i$. Therefore, the starting action yields
\begin{eqnarray}\label{act1}
  S &=& S_{\mbox{\tiny YM}}+S_{\mbox{\tiny MAG}}+S_{\mbox{\tiny diag}}+\int
  d^4x\left(f^{abi}\vartheta^i\occ^ac^b+\frac{\chi}{2}\vartheta^i\vartheta^i\right)\;.
\end{eqnarray}
We define the anomalous dimension $\gamma(g^2)$ of the ghost
operator via
\begin{eqnarray}
\omu\frac{\p}{\p\omu}\left[f^{abi}\occ^ac^b\right]&=&\gamma(g^2)\left[f^{abi}\occ^ac^b\right]=\left(\omu\frac{\p}{\p\omu}\ln
Z_\vartheta\right)\left[f^{abi}\occ^ac^b\right]\;.\end{eqnarray}
From the bare action associated to \eqref{act1}, we deduce
\begin{eqnarray}
  \label{lcoreg}\frac{1}{2}\chi_o\vartheta^i_o\vartheta^i_o&=&\frac{1}{2}\omu^{-\varepsilon}(\chi+\delta\chi)\vartheta^i\vartheta^i\;.
\end{eqnarray}
The so-called LCO parameter $\chi$ is needed to ensure
multiplicative renormalizability: a counterterm $\propto
\vartheta^2$ is needed to kill the divergences in the Green function
$\langle f^{abi}\occ^a(x) c^b(x)f^{abi}\occ^a(y) c^b(y)\rangle$, or
equivalently in the generating functional ${\cal W}(\vartheta)$. It
is clear that divergences $\propto \vartheta^2$ can and do arise. In
principle, $\chi$ is a free parameter. However, as we do not want to
introduce an independent coupling, we shall reexpress $\chi$ in
terms of the gauge coupling $g^2$, in such a way that the
compatibility with the renormalization group is preserved
\cite{Verschelde:1995jj}. We can derive the RG equation for the LCO
parameter $\chi$ from \eqref{lcoreg},
\begin{equation}\label{rgeta}
\omu\frac{\p}{\p\omu}\chi=\left(\beta(g^2)\frac{\p}{\p
g^2}+\gamma_\al(g^2)\al\frac{\p}{\p
\al}\right)\chi=2\gamma(g^2)\chi+\delta(g^2)\;,
\end{equation}
where we defined
\begin{equation}\label{deltarg}
\delta(g^2)=\left(\varepsilon+2\gamma(g^2)-\beta(g^2)\frac{\p}{\p
g^2}-\al\gamma_\al(g^2)\frac{\p}{\p\al}\right)\delta\chi\;.
\end{equation}
Apparently, we require explicit knowledge of $\beta(g^2)$,
$\gamma_\al(g^2)$ and $\delta(g^2)$ before we can fix $\chi(g^2)$ by
solving \eqref{rgeta}. A complete three loop renormalization of QCD
in the MAG in arbitrary colour group has already been carried out in
\cite{Gracey:2005vu}. The only missing information is in fact the RG
function $\delta(g^2)$ as defined in \eqref{deltarg} and the
anomalous dimension $\gamma(g^2)$ of the ghost operator. To deduce
$\delta(g^2)$ we follow the method derived in \cite{Browne:2003uv}.
There the divergences contributing to the counterterm analogous to
$\delta \chi$ were deduced in the massless theory by considering the
corresponding $\vartheta^i$ $2$-point function with no internal
$\vartheta^i$ propagators. As the Feynman graphs are massless and we
are only interested in the divergences, the {\sc Mincer},
\cite{Gorishnii:1989gt,Larin:1991fz}, algorithm written in the
symbolic manipulation language {\sc Form}, \cite{Vermaseren:2000nd},
can be used. The Feynman diagrams are generated automatically using
the {\sc Qgraf} package, \cite{Nogueira:1991ex}, and for our current
problem there are one one loop and twelve two loop Feynman diagrams
to determine. In addition we have also carried out the explicit
renormalization of the operator $f^{abi} \bar{c}^a c^b$ itself at
two loops and verified that the relation derived from the Ward
identities, $\gamma(g^2)$~$=$~$-$~$2\gamma_c(g^2)$ holds, suitably
adapted to our conventions here. This can be regarded as an extra
check on both the algebraic renormalization result as well as the
intricate symbolic manipulation required to derive anomalous
dimensions in the MAG due to the difficulties arising from the split
colour group. See, for instance, \cite{Gracey:2005vu}. Hence, using
the $\MSbar$ renormalization scheme, we obtained the following
results for a general gauge group
\begin{eqnarray}
  \delta(g^2) &=& \delta_0+\delta_1 g^2+\delta_2g^4+\ldots\;,\nonumber \\
\delta_0 &=& -\frac{C_A}{8\pi^2}\;,\delta_1 =
-\frac{1}{2N_A^o\left(16\pi^2\right)^2}\left(N_A^oC_A^2(\al+5)+N_A^dC_A^2(-2\al+22)\right)\,,\nonumber\\
\delta_2&=&
-\frac{1}{32(N_A^o)^2\left(16\pi^2\right)^3}\left((N_A^o)^2(C_A^3(6\al^2+78\al+402)-240C_A^2T_FN_f)\right.\nonumber\\&+&\left.N_A^oN_A^d(C_A^3(60\al^2+96\al\zeta_3+634\al+480\zeta_3+1111)
-608C_A^2T_FN_f)\right.\nonumber\\
&+&\left.(N_A^d)^2(C_A^3(112\al^2-192\al\zeta_3+276\al+2112\zeta_3-1462))\right)\;,
\end{eqnarray}
and
\begin{eqnarray}
  \gamma(g^2) &=& \gamma_0g^2+\gamma_1 g^4+\ldots\;, \nonumber\\
\gamma_0 &=& \frac{1}{2N_A^o\left(16\pi^2\right)}\left(N_A^oC_A(\al+3)+N_A^dC_A(2\al+6)\right)\;,\nonumber\\
\gamma_1 &=&
\frac{1}{48(N_A^o)^2\left(16\pi^2\right)^2}\left((N_A^o)^2(C_A^2(6\al^2+66\al+190)-80C_AT_FN_f)\right.\nonumber\\&+&\left.N_A^oN_A^d(C_A^2(54\al^2+354\al+323)-160C_AT_FN_f)+(N_A^d)^2C_A^2(60\al^2+372\al-510)\right)\;.\nonumber\\
\end{eqnarray}
Using the same notation as \cite{Gracey:2005vu}, $N_A$ is the
dimension of the adjoint representation, whereby $N_A^d$ and $N_A^o$
represent the number of diagonal, respectively off-diagonal,
generators. Of course, $N_A^d+N_A^o=N_A$. $N_f$ is the number of
quark flavours, while $T_F$ and $C_A$ are Casimir operators.
Specifying to $SU(N)$, one has $N_A^d=N-1$, $N_A^o=N(N-1)$,
$T_F=\half$ and $C_A=N$.

For simplicity, we shall only determine the potential in the case of
$SU(2)$ as gauge group without flavours. If $N=2$, there is only one
ghost condensate, as $SU(2)$ has only one $U(1)$ subgroup. In that
case, we have
\begin{eqnarray}
\delta_0 &=& -\frac{1}{4\pi^2}\,, \delta_1=
-\frac{32}{(16\pi^2)^2}\,,\nonumber\\a_0& =&
\left(-2\al+\frac{8}{3}-\frac{6}{\al}\right)\frac{1}{16\pi^2}\,,a_1
=
\frac{1}{3}\left(-12\al^2-156\al+52+\frac{20}{\al}\right)\frac{1}{(16\pi^2)^2}\;,
\end{eqnarray}
while
\begin{eqnarray}\label{a2feynrge15}
    \beta(g^2)&=&-\varepsilon g^2-2\left(\beta_0 g^4+\beta_1
g^6\right)+\ldots\;,\nonumber\\
\beta_0&=&\frac{22}{3}\frac{1}{16\pi^2} \;,
\beta_1=\frac{136}{3}\frac{1}{\left(16\pi^2\right)^2}\;.
\end{eqnarray}
Equation \eqref{rgeta} can be solved by making $\chi$ a Laurent
series in $g^2$,
\begin{equation}
    \chi(g^2,\al)=\frac{\chi_0(\al)}{g^2}+\chi_1(\al)+\ldots\;.
\end{equation}
Substituting this in \eqref{rgeta}, we obtain the following
differential equations in $\al$ for the first two coefficients
$\chi_0$ and $\chi_1$.
\begin{eqnarray}
\label{eta0}2\beta_0\chi_0+\al
a_0\frac{\partial\chi_0}{\partial\al}&=&2\gamma_0\chi_0+\delta_0\;,\\
\label{eta1}2\beta_1\chi_0+\al
a_0\frac{\partial\chi_1}{\partial\al}+\al
a_1\frac{\partial\chi_0}{\partial\al}&=&2\gamma_0\chi_1+2\gamma_1\chi_0
+\delta_1\;.
\end{eqnarray}
Solving yields
\begin{equation}\label{chi0}
    \chi_0=\frac{6\al+C_0}{3\al^2-4\al+9}\;.
\end{equation}
For $\chi_1$, we have not been able to find a closed expression. An
integral representation is given by
\begin{eqnarray}\label{chi1}
    \chi_1&=&\frac{e^{-\frac{22}{\sqrt{23}}\mbox{\tiny
    ArcTan}\left(\frac{3\al}{\sqrt{23}}-\frac{2}{\sqrt{23}}\right)}}{3\al^2-4\al+9}\nonumber\\
    &&\times\int_{C_1}^\al\left[\vphantom{\frac{e^{\frac{22}{\sqrt{23}}\mbox{\tiny
    ArcTan}\left(\frac{3x}{\sqrt{23}}-\frac{2}{\sqrt{23}}\right)}}{\pi^2(3x^2-4x+9)^2}}\left(\frac{567}{4}x-468x^2+\frac{95}{2}C_0-\frac{129}{2}C_0x+15C_0x^3+54x^3+\frac{153}{4}x^4-\frac{27}{4}x^5+\frac{1107}{4}\right)\right.\nonumber\\
    &&\left.\qquad\times\frac{e^{\frac{22}{\sqrt{23}}\mbox{\tiny
    ArcTan}\left(\frac{3x}{\sqrt{23}}-\frac{2}{\sqrt{23}}\right)}}{\pi^2(3x^2-4x+9)^2}\right]dx\;.
\end{eqnarray}
$C_0$ and $C_1$ are constants of integration.

Let us recall that the exact vacuum energy itself will not depend on
the choice of the gauge parameter $\al$, which can be proven
completely similarly as we already did before in
\cite{Dudal:2003by,Dudal:2004rx}. We also recall that we introduced
a method to circumvent the gauge parameter dependence of the
explicitly calculated $\Evac$, caused by the fact that we are forced
to work at a finite order, so that we never obtain that
$J=0\;exactly$. Essentially, we introduced a ``compensating'' gauge
dependent function that was determined to remove the gauge
dependence. If we introduce the following unity
\begin{equation}\label{eenheid}
    1=\mathcal{N}\int \mathcal D\sigma e^{-\frac{i}{2\chi}\int d^4x\left(\frac{\sigma^i}{g}-\chi \theta^i-f^{abi}\occ^a
    c^b\right)^2}\;,
\end{equation}
with $\mathcal{N}$ the appropriate normalization, we are led to the
following action
\begin{equation}\label{nieuweactie}
    S^\prime=S_{\mbox{\tiny YM}}+S_{\mbox{\tiny MAG}}+S_{\mbox{\tiny diag}}+\int d^4x\left(-\frac{\sigma^i\sigma^i}{2g^2\chi}+\frac{1}{\chi g}\sigma^i f^{abi}\occ^a c^b-\frac{1}{2\chi}f^{abi}f^{cdi}\occ^a c^b\occ^c
    c^d+\theta^i\frac{\sigma^i}{g}\right)\;,
\end{equation}
with the identification
\begin{equation}\label{ident}
    \langle gf^{abi}\occ^a c^b\rangle=\langle\sigma^i\rangle\;.
\end{equation}
Following the analysis of \cite{Dudal:2003by,Dudal:2004rx}, one can
show that the vacuum energy, given in terms of the effective
potential $V(\sigma)$ as
\begin{equation}\label{ident2}
    \Evac=V(\sigma_*)\;,\qquad\mbox{with}\;\sigma_*
    \textrm{ the global minimum of } V(\sigma)\;,
\end{equation}
shall formally not depend on the gauge parameter, making use of the
BRST symmetry which can be extended naturally to the extra field by
means of
\begin{equation}\label{brstnieuw}
    s\sigma^i=s(gf^{abi}\occ^ac^b)=gf^{abi}b^ac^b-g^2f^{abi}f^{bci}\occ^a c^c
    c^i-\frac{g^2}{2}f^{abi}f^{bcd}\occ^a c^c c^d\;.
\end{equation}
We shall not repeat the proof here, as it would be merely a
notational adaptation of the analogous results in
\cite{Dudal:2003by,Dudal:2004rx}. We emphasize the use of the word
\emph{formally}, as we are forced to work at a finite order. The
gauge parameter independence proof is only valid when we would work
to all orders. The problem relies on the fact that an important step
in the quoted proof is that the sources $\vartheta^i$ become zero
when the gap equation leading to the minimum of the effective
potential is solved. However, as the effective potential $V(\sigma)$
itself shall only be calculated in a loop expansion, we shall only
have $\vartheta^i=0$ up to a certain order, because $\vartheta^i\sim
\frac{\p V(\sigma)}{\p\sigma^i}$. Consequently, at finite order,
residual $\al$-dependence will slip into the final expression for
the vacuum energy. To cure the $\alpha$-dependence at finite order
precision, we shall rely on the formalism developed
\cite{Dudal:2003by,Dudal:2004rx}. We apply a transformation to the
fields and the sources,
\begin{equation}\label{v1}
    \sigma^i=\frac{\widetilde{\sigma}}{\mathcal{F}(g^2,\al)}\;,\qquad\qquad
    \vartheta^i=\widetilde{\vartheta}^i\mathcal{F}(g^2,\al)\;,
\end{equation}
with
\begin{equation}
{\cal F}(g^2,\al)=1+f_0(\al)g^2+f_1(\al)g^4+\ldots\;,
\end{equation}
to arrive at the following action
\begin{eqnarray}\label{nieuweactie2}
    S^\prime=S_{\mbox{\tiny YM}}+S_{\mbox{\tiny MAG}}+S_{\mbox{\tiny diag}}&+&\int d^4x\left(-\frac{\wsigma^i\wsigma^i}{2g^2\mathcal{F}(g^2,\al)\chi}+\frac{1}{g\chi\mathcal{F}(g^2,\al) }\wsigma^i f^{abi}\occ^a c^b\right.\nonumber\\&&\left.-\frac{1}{2\chi}f^{abi}f^{cdi}\occ^a c^b\occ^c
    c^d+\widetilde{\vartheta}^i\frac{\wsigma^i}{g}\right)\;.
\end{eqnarray}
In the $SU(2)$ case, in which case there is only one field
$\sigma\equiv\sigma^3$, the tree level off-diagonal ghost propagator
will read
\begin{equation}\label{offdiagghost}
    \langle \occ^a
    c^b\rangle_q=i\frac{-\delta^{ab}q^2+v\epsilon^{ab}}{q^4+v^2}\;,
\end{equation}
where we set
\begin{equation}\label{v}
    v=\frac{g}{\chi_0}\langle\wsigma\rangle\;.
\end{equation}
We notice that the actions \eqref{nieuweactie} and
\eqref{nieuweactie2} are \emph{exactly} equivalent as they are
connected via the transformations \eqref{v1}, however when working
up to a certain order the coefficient functions $f_i(\al)$ can enter
the results. We shall precisely use these to enforce the gauge
parameter independence of the vacuum energy. We shall demand that
\begin{equation}\label{requir}
    \frac{d\Evac}{d\al}=0\Rightarrow \mbox{first order differential equations in $\al$ for
    $f_i(\al)$}\;.
\end{equation}
As an initial condition for the vacuum energy, we shall use the
Landau gauge result. In \cite{Dudal:2003dp}, we analyzed the ghost
condensate $\langle f^{ABC}\occ^B c^C\rangle$ in the Landau gauge.
By connecting the MAG with the Landau gauge in \cite{Dudal:2004rx},
we argued that we can use the Landau gauge as the ``initial
condition gauge'' to match the vacuum energy of any other gauge to
that of the Landau gauge, given that the other gauge can be linked
to the Landau gauge in a renormalizable fashion. Of course, we
should also find a renormalizable interpolating ghost operator. In
the present case, it is given by the expression
\begin{equation}\label{int2}
    \int d^4x\left(\vartheta^i f^{abi}\occ^a c^b+\kappa\vartheta^c \left(f^{abc}\occ^a
    c^b+f^{jbc}\occ^j
    c^b+f^{ajc}\occ^a
    c^j\right)+\chi \vartheta^i\vartheta^i +
    \kappa\chi'\vartheta^c\vartheta^c\right)\;,
\end{equation}
where $k$ is an interpolating parameter\footnote{We recall that
$\al,\kappa$ are gauge parameters, such that the MAG corresponds
to $\kappa=0$ while the Landau gauge to $(\al=0,\kappa=1)$
\cite{Dudal:2004rx}.} and $\vartheta^i,\vartheta^c$ external
sources. The Landau gauge vacuum energy was established to be
\begin{equation}\label{potnew5}
    \Evac^{\mbox{\tiny
    Landau}}=-\frac{1}{32\pi^2}e^{\frac{56}{33}}\lms^4\approx-0.017\lms^4\;.
\end{equation}
The $SU(2)$ MAG effective action reads at one loop, again using the
$\MSbar$ scheme,
\begin{equation}\label{potmag2}
    V_{1}(\sigma)=\frac{\sigma^2}{2\chi_0}\left(1-\frac{\chi_1}{\chi_0}g^2\right)+\frac{1}{32\pi^2}\frac{g^2\sigma^2}{\chi_0^2}\left(\ln\frac{g^2\sigma^2}{\chi_0^2\omu^4}-3\right)\;.
\end{equation}
Performing the transformation yields the potential
\begin{equation}\label{potmag3}
    V_{1}(\wsigma)=\frac{\wsigma^2}{2\chi_0}\left(1-\left(\frac{\chi_1}{\chi_0}+2f_0\right)g^2\right)+\frac{1}{32\pi^2}\frac{g^2\wsigma^2}{\chi_0^2}\left(\ln\frac{g^2\wsigma^2}{\chi_0^2\omu^4}-3\right)\;.
\end{equation}
The gap equation $\frac{dV_1}{d\wsigma}=0$ leads to
\begin{equation}\label{potmag4}
    \frac{\wsigma}{\chi_0}\left(1-\left(\frac{\chi_1}{\chi_0}+2f_0\right)g^2\right)+\frac{1}{16\pi^2}\frac{g^2\wsigma}{\chi_0^2}\left(\ln\frac{g^2\wsigma^2}{\chi_0^2\omu^4}-3\right)+\frac{1}{16\pi^2}\frac{g^2\wsigma}{\chi_0^2}=0\;.
\end{equation}
Assuming that $v_*$ is a solution of the previous equation written
in terms of the variable $v$ as defined in \eqref{v}, we obtain as
vacuum energy
\begin{equation}\label{potmag5}
\Evac^{\mbox{\tiny MAG}}=-\frac{1}{32\pi^2}v_*^2\;.
\end{equation}
Now, by construction of the method, the functions $f_i(\al)$ are
fixed to ensure that $\Evac^{\mbox{\tiny Landau}}=\Evac^{\mbox{\tiny
MAG}}$. Doing so, we can in fact solve
\begin{equation}\label{potmag6}
-\frac{1}{32\pi^2}v_*^2=-\frac{1}{32\pi^2}e^{\frac{56}{33}}\lms^4\;,
\end{equation}
or
\begin{equation}\label{potmag7}
v_*=e^{\frac{28}{33}}\lms^2\approx 2.34\lms^2\;.
\end{equation}
For comparison, the lattice group of \cite{Mendes:2006kc} quote a
(preliminary) estimate of $v\approx 1.3\, \mbox{GeV}^2$. Using
$\lms\approx 275\, \mbox{MeV}$ in the case of $SU(2)$
\cite{Michael:1992nj}, we find
\begin{equation}\label{potmag8}
v_*\approx 0.18 \,\mbox{GeV}^2\;.
\end{equation}
It is instructive to have a look at the effective coupling constant.
Assuming that we solve the gap equation at a scale $\omu^2=v_*^2$ in
order to kill large logarithms and using the one loop result
\eqref{potmag4}, we deduce that, for any $\alpha$,
\begin{equation}\label{sol}
\left.\frac{g^2N}{16\pi^2}\right\vert_{N=2}=\frac{9}{28}\;,
\end{equation}
which is sufficiently small to speak about at least qualitatively
acceptable results. We notice that our value is considerable smaller
than the lattice value. However, continuum effects should still be
investigated on the lattice, while we employed perturbation theory
at lowest order. One can imagine other sources of nonperturbative
effects that contributes to the condensate. Anyhow, analytical
continuum calculations as well as lattice simulations seem to favour
a nonvanishing ghost condensate.

We also see that we do not need explicit knowledge of the
$f_i(\al)$-terms to obtain the desired results, by taking into
account how the functions $f_i(\al)$ are fixed. For completeness,
one could determine $f_0(\al)$ by matching the solution of the gap
equation \eqref{potmag4} at $\omu^2=v_*$, being
\begin{equation}\label{sol2}
\left.\frac{g^2N}{16\pi^2}\right\vert_{N=2}=\frac{1-\frac{1}{8\pi^2\chi_0}}{8\pi^2\left(\frac{\chi_1}{\chi_0}+2f_0\right)}\;,
\end{equation}
with the already fixed solution \eqref{sol}, so there is absolutely
no need to solve the defining differential equations \eqref{requir}.

\section{Consequences of the ghost condensate}
First of all, there is the obvious breaking of the $\delta$-symmetry
\eqref{symgen} since $\langle f^{abi}\occ^a c^b\rangle=\half\langle
\delta(f^{abi}\occ^a\occ^b)\rangle$. Let us have a look at the
associated current in the $SU(2)$ case under study\footnote{A little
more cumbersome calculations will lead to the same conclusion in the
general $SU(N)$ case.}, while setting $\alpha=0$ in \eqref{totaal}
to immediately recover the MAG \eqref{MAG2}. Doing so, the symmetry
generator $\delta$, defined in \eqref{symgen}, is given
by\footnote{$\epsilon^{12}=-\epsilon^{21}=1$,
$\epsilon^{11}=\epsilon^{22}=0$.}
\begin{equation}\label{symgen2}
    \delta=\int d^4x\left(c^a\frac{\delta}{\delta \occ^a}+g \epsilon^{ab} c^b c\frac{\delta}{\delta
    b^a}\right)
\end{equation}
in the functional form, so that we find after its application on the
action \eqref{totaal},
\begin{eqnarray}\label{current}
    &&0=\delta S= \int d^4x \delta \mathcal{L}\nonumber\\
    &\Rightarrow& \left(c^a\frac{\delta S}{\delta \occ^a}+g \epsilon^{ab} c^b c\frac{\delta S}{\delta b^a}
    \right)=\p_\mu (c^a D_\mu^{ab} c^b)\;.
\end{eqnarray}
After using the equations of motion, it follows that
\begin{eqnarray}\label{currentbis}
    \mathcal{K}_\mu=c^a D_\mu^{ab} c^b
\end{eqnarray}
is the associated conserved current, $\p_\mu\mathcal{K}_\mu=0$. Now,
it turns out that this current $\mathcal{K}_\mu$ can be brought into
the following useful form
\begin{equation}\label{cur1}
    \mathcal{K}_\mu=s(A_\mu^ac^a)\;,
\end{equation}
which can be checked by using the definition of the BRST
transformation $s$, as given in \eqref{BRST}.

We would like to point out here that the ghost condensate
$\left\langle \epsilon^{ab}\occ^a c^b\right\rangle$ does not break
the BRST symmetry, since $s (\epsilon^{ab}\occ^a c^b)\neq0$, so that
due to the nilpotency, we certainly do have $\epsilon^{ab}\occ^a
c^b\neq s(\ldots)$, meaning that the ghost condensate is not an
order parameter for the BRST symmetry. In order to avoid confusion,
let us mention that, in the case that one would study the
condensates $\langle \epsilon^{ab}\occ^a \occ^b\rangle$ and $\langle
\epsilon^{ab} c^a c^b\rangle $, corresponding to an equivalent
vacuum, then a nilpotent BRST charge also exists. We refer to
\cite{Dudal:2003dp} for a detailed discussion of the completely
analogous arguments in the Landau gauge.

As we have already mentioned, there exists another version of the
MAG \cite{Dudal:2002ye}, in which case the diagonal gauge fixing
also respects the anti-BRST symmetry $\overline{s}$. The ghost
condensate discussed here then breaks this anti-BRST symmetry. We
did choose the diagonal gauge fixing \eqref{diag}, as this
corresponds to the Landau gauge for the diagonal sector, a gauge
also used in the lattice simulations corresponding to the MAG
\cite{Amemiya:1998jz,Bornyakov:2003ee,Mendes:2006kc}.

Returning to the current $\mathcal{K}_\mu$ written down in
\eqref{cur1}, we can use the fact that it
 is BRST exact. As a consequence, the Goldstone
boson associated with the broken $\delta$-invariance will cancel
from the physical spectrum, as it will belong to a BRST exact state,
and physical states are defined as BRST invariant states, modulo the
(trivially) invariant exact states. We rely here on the fact that
the current corresponding to a spontaneously broken symmetry stands
in a direct correspondence with the associated Goldstone boson
\cite{Peskin:1995ev}, namely the current can be used to
create/annihilate the Goldstone boson. Inserting this current into a
physical gauge invariant, and thus a fortiori BRST invariant,
correlator then immediately leads to a vanishing result.

A nonvanishing ghost condensate also strongly influences the
behaviour of the off-diagonal ghost propagator,
\eqref{offdiagghost}. We notice the safe infrared behaviour when
$p^2\to 0$. However, the ghost condensate will also enter the
off-diagonal gluon propagator through radiative corrections. At one
loop order, the quartic $A^2\occ c$ coupling will give rise to an
effective 1PI off-diagonal gluon mass $\delta M^2$ given by
\cite{Dudal:2002xe}
\begin{equation}\label{mass1}
    \delta M^2= -\frac{g^2}{16\pi^2}v_*<0\;.
\end{equation}
Clearly, this would be a tachyonic off-diagonal gluon mass,
indicative of instabilities in the ghost condensed vacuum. There is
however a resolution to this problem.

So far we have only considered the contribution of the ghost
condensate to an effective gluon mass. If we assume that we would
have started with a sufficiently large positive tree level
off-diagonal gluon mass squared, the \emph{loop effect}
\eqref{mass1} should merely introduce a shift in the tree level
value, together with potential other shifts coming from the other
interactions.

We have investigated the dynamical generation of a off-diagonal
gluon mass $m^2_{od}$ with similar techniques as employed here
\cite{Dudal:2004rx}. We considered the dimension 2 operator
$\frac{1}{2}A_\mu^a A_\mu^a+\al\occ^a c^a$, which is on-shell BRST
invariant as encoded in the integrated $\lambda$-equation
\eqref{lambda-eq}, and successfully constructed the effective
potential at one loop in the $SU(2)$ case, leading to a finite value
\begin{equation}\label{mass2}
    m_{od}^2=\sqrt{\frac{3}{2}e^{\frac{17}{6}}}\lms^2\approx
    5.05\lms^2\;,
\end{equation}
in the MAG limit $\al\to0$. In a meaningful perturbative expansion,
one should certainly have $\frac{g^2N}{16\pi^2}<1$, so that upon
comparing the numbers \eqref{potmag7}, \eqref{sol}, \eqref{mass1}
and \eqref{mass2}, we conclude that the ghost condensation will
induce a negative shift in the mass \eqref{mass2}, however the net
result will be still positive. Further one loop corrections will
come from the pure gluonic vacuum polarization. A complete account
of similar effects in the Landau gauge was presented in
\cite{Capri:2005vw}.

Let us finally mention that the diagonal sector will not be
influenced by the ghost nor gluon condensate, since the $U(1)^{N-1}$
Ward identity \eqref{U(1)-eq} forbids a diagonal gluon mass, while
the diagonal antighost equation \eqref{anti-gh-eq} excludes a
diagonal ghost mass.

\section{Conclusion}
We have given evidence for the existence of a mass dimension 2 ghost
condensate in the MAG. We used the LCO formalism
\cite{Verschelde:1995jj} to construct a sensible effective potential
for the ghost operator $f^{abi}\occ^ac^b$. We proved the
renormalizability to all orders of perturbation theory, and
explicitly calculated the one loop effective potential, thereby
finding a nonvanishing ghost condensate since $\langle
f^{abi}\occ^ac^b\rangle$ corresponds to a vacuum with lower energy.
The mere existence of a ghost condensate is in qualitative agreement
with recent lattice data \cite{Cucchieri:2005yr,Mendes:2006kc}.

There are a few interesting open questions related to the ghost
condensate that deserves further investigation. Since $\langle
f^{abi}\occ^ac^b\rangle$ serves as an order parameter for a
symmetry, it should be investigated whether this symmetry might get
restored if we would allow for finite temperature effects.

In \cite{Capri:2005tj,Capri:2006cz}, it was discussed how a
(partial) treatment of Gribov copy effects in the MAG might be
handled via a restriction of the domain of path integration along
the lines of Gribov's original approach \cite{Gribov:1977wm}. Since
this restriction also seriously alters the infrared behaviour of the
propagators, it would be instructive to find out whether there is a
significant change in the obtained values of the ghost condensate.

So far, the mass generating mechanism in the Landau gauge or MAG was
in fact depending on the gauge, since the used operator is not gauge
invariant, nevertheless the qualitative features of the analytical
results in the MAG \cite{Dudal:2004rx} are in quite good agreement
with lattice data \cite{Amemiya:1998jz,Bornyakov:2003ee}. In
principle, $A^2$ is gauge invariant when used in the Landau gauge,
as it is formally equivalent to the gauge invariant functional
$A^2_{\min}$, obtained by taking the absolute minimum of $A^2$ along
its gauge orbit. However, outside of the Landau gauge, it is unclear
how to use this operator. In
\cite{Capri:2005dy,Capri:2006ne,Dudal:2007ch}, we developed a local,
renormalizable non-Abelian gauge invariant action, based on the
nonlocal mass operator $F\frac{1}{D^2}F$, which could serve as a
starting point to discuss a gauge invariant mechanism behind
dynamical mass parameters in e.g. the gluon propagator (or physical
correlators). Due to the gauge invariance, we therefore expect the
\emph{same} tree level mass in the diagonal and off-diagonal sector,
even in the MAG. The question arises what might cause the possible
difference between the diagonal and off-diagonal sector? A possible
explanation might be the ghost condensate, in a fashion similar to
what we studied in \cite{Capri:2005vw}. Said otherwise, the ghost
condensate could play an important role clarifying the mechanism(s)
behind Abelian dominance.

\section*{Acknowledgments.}
The Conselho Nacional de Desenvolvimento Cient\'{\i}fico e
Tecnol\'{o}gico
(CNPq-Brazil), the SR2-UERJ and the Coordena{\c{c}}{\~{a}}o de Aperfei{\c{c}}%
oamento de Pessoal de N{\'\i}vel Superior (CAPES) are gratefully
acknowledged for financial support. D.~Dudal is supported by the
``Special Research Fund'' of Ghent University.


\begin{thebibliography}{99}
\bibitem{Peskin:1995ev}
 M.~E.~Peskin and D.~V.~Schroeder, \emph{An Introduction To Quantum Field
Theory}, Addison-Wesley Publishing Company (1995).

\bibitem{Henneaux:1992ig}
 M.~Henneaux and C.~Teitelboim, \emph{Quantization of gauge systems}, Princeton University Press (1992).

\bibitem{Piguet:1995er}
O.~Piguet and S.~P.~Sorella, Lect.\ Notes Phys.\  {\bf M28} (1995)
1.

\bibitem{Gribov:1977wm}
V.~N.~Gribov, Nucl.\ Phys.\  B {\bf 139} (1978) 1.

\bibitem{Gracey:2006dr}
J.~A.~Gracey, JHEP {\bf 0605} (2006) 052.

\bibitem{Dudal:2005na}
D.~Dudal, R.~F.~Sobreiro, S.~P.~Sorella and H.~Verschelde, Phys.\
Rev.\  D {\bf 72} (2005) 014016.

\bibitem{Kugo:1979gm}
T.~Kugo and I.~Ojima, Prog.\ Theor.\ Phys.\ Suppl.\  {\bf 66} (1979)
1.

\bibitem{Kugo:1995km}
T.~Kugo, hep-th/9511033.

\bibitem{Maas:2006qw}
A.~Maas, A.~Cucchieri and T.~Mendes, Braz.\ J.\ Phys.\  {\bf 37}
(2007) 219.

\bibitem{Watson:2001yv}
P.~Watson and R.~Alkofer, Phys.\ Rev.\ Lett.\  {\bf 86} (2001) 5239.

\bibitem{Lerche:2002ep}
C.~Lerche and L.~von Smekal, Phys.\ Rev.\  D {\bf 65} (2002) 125006.

\bibitem{Zwanziger:2001kw}
D.~Zwanziger, Phys.\ Rev.\  D {\bf 65} (2002) 094039.

\bibitem{Schaden:1999ew}
M.~Schaden, hep-th/9909011.

\bibitem{Kondo:2000ey}
K.~I.~Kondo and T.~Shinohara, Phys.\ Lett.\  B {\bf 491} (2000) 263.

\bibitem{'tHooft:1981ht}
G.~'t Hooft, Nucl.\ Phys.\  B {\bf 190} (1981) 455.

\bibitem{Min:1985bx}
H.~Min, T.~Lee and P.~Y.~Pac, Phys.\ Rev.\  D {\bf 32} (1985) 440.

\bibitem{Fazio:2001rm}
A.~R.~Fazio, V.~E.~R.~Lemes, M.~S.~Sarandy and S.~P.~Sorella, Phys.\
Rev.\  D {\bf 64} (2001) 085003.

\bibitem{Dudal:2002xe}
D.~Dudal and H.~Verschelde, J.\ Phys.\ A  {\bf 36} (2003) 8507.

\bibitem{Verschelde:1995jj}
H.~Verschelde, Phys.\ Lett.\  B {\bf 351} (1995) 242.

\bibitem{Dudal:2003by}
D.~Dudal, H.~Verschelde, J.~A.~Gracey, V.~E.~R.~Lemes,
M.~S.~Sarandy, R.~F.~Sobreiro and S.~P.~Sorella,  JHEP {\bf 0401}
(2004) 044.

\bibitem{Dudal:2004rx}
D.~Dudal, J.~A.~Gracey, V.~E.~R.~Lemes, M.~S.~Sarandy,
R.~F.~Sobreiro, S.~P.~Sorella and H.~Verschelde, Phys.\ Rev.\  D
{\bf 70} (2004) 114038.

\bibitem{Ezawa:1982bf}
Z.~F.~Ezawa and A.~Iwazaki, Phys.\ Rev.\  D {\bf 25} (1982) 2681.

\bibitem{Kondo:2001nq}
K.~I.~Kondo, Phys.\ Lett.\  B {\bf 514} (2001)
335.

\bibitem{Amemiya:1998jz}
K.~Amemiya and H.~Suganuma, Phys.\ Rev.\  D {\bf 60} (1999) 114509.

\bibitem{Bornyakov:2003ee}
V.~G.~Bornyakov, M.~N.~Chernodub, F.~V.~Gubarev, S.~M.~Morozov and
M.~I.~Polikarpov, Phys.\ Lett.\  B {\bf 559} (2003) 214.

\bibitem{Capri:2005vw}
M.~A.~L.~Capri, D.~Dudal, J.~A.~Gracey, V.~E.~R.~Lemes,
R.~F.~Sobreiro, S.~P.~Sorella and H.~Verschelde, Phys.\ Rev.\  D
{\bf 73} (2006) 014001.

\bibitem{Suzuki:2004dw}
T.~Suzuki, K.~Ishiguro, Y.~Mori and T.~Sekido, Phys.\ Rev.\ Lett.\
{\bf 94} (2005) 132001.

\bibitem{Curci:1976bt}
G.~Curci and R.~Ferrari, Nuovo Cim.\  A {\bf 32} (1976) 151.

\bibitem{Dudal:2002ye}
D.~Dudal, V.~E.~R.~Lemes, M.~S.~Sarandy, S.~P.~Sorella and
M.~Picariello,
  JHEP {\bf 0212} (2002) 008.

\bibitem{Lemes:2002jv}
V.~E.~R.~Lemes, M.~S.~Sarandy, S.~P.~Sorella, M.~Picariello and
A.~R.~Fazio, Mod.\ Phys.\ Lett.\  A {\bf 18} (2003) 711.

\bibitem{Lemes:2002rc}
V.~E.~R.~Lemes, M.~S.~Sarandy and S.~P.~Sorella, Annals Phys.\  {\bf
308} (2003) 1.

\bibitem{Dudal:2003dp}
D.~Dudal, H.~Verschelde, V.~E.~R.~Lemes, M.~S.~Sarandy, S.~P.~
Sorella, M.~Picariello, A.~Vicini and J.~A.~Gracey, JHEP {\bf 0306}
(2003) 003.

\bibitem{Cucchieri:2005yr}
A.~Cucchieri, T.~Mendes and A.~Mihara, Phys.\ Rev.\  D {\bf 72}
(2005) 094505.

\bibitem{Mendes:2006kc}
T.~Mendes, A.~Cucchieri and A.~Mihara,  AIP Conf.\ Proc.\  {\bf 892}
(2007) 203.

\bibitem{Gracey:2005vu} J.~A.~Gracey, JHEP {\bf 0504} (2005)
012.

\bibitem{Browne:2003uv}
R.~E.~Browne and J.~A.~Gracey, JHEP {\bf 0311} (2003) 029.

\bibitem{Gorishnii:1989gt}
S.~G.~Gorishnii, S.~A.~Larin, L.~R.~Surguladze and F.~V.~Tkachov,
Comput.\ Phys.\ Commun.\  {\bf 55} (1989) 381.

\bibitem{Larin:1991fz}
S.~A.~Larin, F.~V.~Tkachov and J.~A.~M.~Vermaseren, \emph{The Form
Version Of Mincer}, NIKHEF-H-91-18.

\bibitem{Vermaseren:2000nd}
J.~A.~M.~Vermaseren, math-ph/0010025.

\bibitem{Nogueira:1991ex}
P.~Nogueira, J.\ Comput.\ Phys.\  {\bf 105} (1993) 279.

\bibitem{Michael:1992nj}
C.~Michael, Phys.\ Lett.\ B {\bf 283} (1992) 103.

\bibitem{Capri:2005tj}
M.~A.~L.~Capri, V.~E.~R.~Lemes, R.~F.~Sobreiro, S.~P.~Sorella and
R.~Thibes, Phys.\ Rev.\  D {\bf 72} (2005) 085021.

\bibitem{Capri:2006cz}
M.~A.~L.~Capri, V.~E.~R.~Lemes, R.~F.~Sobreiro, S.~P.~Sorella and
R.~Thibes, Phys.\ Rev.\  D {\bf 74} (2006) 105007.

\bibitem{Capri:2005dy}
M.~A.~L.~Capri, D.~Dudal, J.~A.~Gracey, V.~E.~R.~Lemes,
R.~F.~Sobreiro, S.~P.~Sorella and H.~Verschelde, Phys.\ Rev.\  D
{\bf 72} (2005) 105016.

\bibitem{Capri:2006ne}
M.~A.~L.~Capri, D.~Dudal, J.~A.~Gracey, V.~E.~R.~Lemes,
R.~F.~Sobreiro, S.~P.~Sorella and H.~Verschelde, Phys.\ Rev.\  D
{\bf 74} (2006) 045008.

\bibitem{Dudal:2007ch}
D.~Dudal, N.~Vandersickel and H.~Verschelde, Phys.\ Rev.\  D {\bf
76} (2007) 025006.

\end{thebibliography}
\end{document}